# Pair-breaking quantum phase transition in superconducting nanowires

Hyunjeong Kim[1], Fédéric Gay[2], Adrian Del Maestro[3], Benjamin Sacépé[2], and Andrey Rogachev[1]

[1]Department of Physics and Astronomy, University of Utah, Salt Lake City, Utah 84112, USA
[2]Univ. Grenoble Alpes, CNRS, Grenoble INP, Institut Néel, 38000 Grenoble, France
[3]Department of Physics, University of Vermont, Burlington, Vermont 05405, USA

**A quantum phase transition (QPT) between distinct ground states of matter is a wide-spread phenomenon in nature, yet there are only a few experimentally accessible systems where the *microscopic* mechanism of the transition can be tested and understood. These cases are unique and form the experimentally established foundation for our understanding of quantum critical phenomena. Here we report the discovery that a magnetic-field-driven QPT in superconducting nanowires – a prototypical 1d-system – can be fully explained by the critical theory of pair-breaking transitions characterized by a correlation length exponent $\nu \approx 1$ and dynamic critical exponent $z \approx 2$. We find that in the quantum critical regime, the electrical conductivity is in agreement with a theoretically predicted scaling function and, moreover, that the theory *quantitatively* describes the dependence of conductivity on the critical temperature, field magnitude and orientation, nanowire cross sectional area, and microscopic parameters of the nanowire material. At the critical field, the conductivity follows a $T^{(d-2)/z}$ dependence predicted by phenomenological scaling theories and more recently obtained within a holographic framework. Our work uncovers the microscopic processes governing the transition: The pair-breaking effect of the magnetic field on interacting Cooper pairs overdamped by their coupling to electronic degrees of freedom. It also reveals the universal character of continuous quantum phase transitions.**

Quantum phase transitions occur in many systems including magnetic materials,[1] superconductors,[2,3,4,5] cold atomic gases[6] and also atomic nuclei[7] and stars.[8] Similar to the thermal fluctuations in classical temperature-driven phase transitions, strong quantum fluctuations near the critical point of a QPT lead to the emergence of universal long-range behavior, which can be common in very diverse systems. However, for a complete description of a QPT one must also identify and quantitatively incorporate into a theory specific microscopic processes which drive a system across the critical point and induce the fluctuations. Examples where such complete theories can be experimentally tested are scarce and include the 2-channel Kondo effect in quantum dots[9] and Luttinger liquid behavior in materials composed of weakly coupled 1-dimensional (1d) spin-chains.[10]

Superconducting systems present special interest in the context of QPT because the fluctuations near the critical point can lead to the formation unconventional superconducting phases (most notably this is one the scenarios for high temperature superconductivity in the cuprates[11]). They also present a challenge – despite many years of efforts and overall success of phenomenological finite-size scaling analyses,[12,13,14] the microscopic mechanism of QPTs in 2d superconductors is still debated. In contrast, the physics of a QPT becomes much more transparent in 1d superconductors. Here we show that essentially all long-range and microscopic characteristics of QPT driven in superconducting nanowires by magnetic field can be described by a pair-breaking critical theory.

A 1d superconductor can be defined as a wire with diameter smaller than $\pi 2^{1/2}\xi(0)$, where $\xi(0)$ is the zero-temperature Ginzburg-Landau coherence length.[15] This condition ensures that vortices do not form within the wire and that the superconducting order parameter is approximately constant at a given cross-section. 1d superconductors, as all 1d systems, are strongly affected by fluctuations, which can be both thermally activated or caused by quantum tunneling.[16,17,18] The rate of fluctuations increases exponentially in thin wires.

Many experimental studies[15,16,19,20] have shown that reducing the nanowire diameter can drive a 1d superconductor into an insulating state. However, the microscopic mechanism of this process and the nature of the insulating phase (Bose insulator, Fermi insulator or some other state of matter) remain unclear. Better understood is the case when superconductivity is destroyed by a magnetic field.[19,21] Figure 1 shows an expected phase diagram for this process

The field acts on orbital and spin degrees of freedom of a Cooper pair as a *pair breaker*, cutting off the logarithmic divergence in the pairing susceptibility and setting a critical field $B_c$ above which no bulk superconductivity is possible, even at $T = 0$. At finite temperature, both the amplitude of the superconducting order parameter and the critical temperature are suppressed, as shown in the figure by the dashed line. Also, close to, but below $B_c$, the superconducting gap in nanowires shrinks to zero and the superconductivity becomes gapless; in this regime the superconducting condensate co-exists with normal quasiparticles. The state of the wire above $B_c$ is, within the temperature range of our experiments, a normal disordered metal which experiences pairing fluctuations near the Fermi surface as a result of its proximity to the superconducting state. This can be pictured as a temporal conversion of a section of a wire into the superconducting state that leads to a measurable enhancement of the conductivity. Such corrections, due to both quantum and thermal fluctuations, were computed by Shah and Lopatin using a perturbative diagrammatic formalism.[22] The part of the phase diagram described by this theory is schematically shown in Fig. 1 in green.

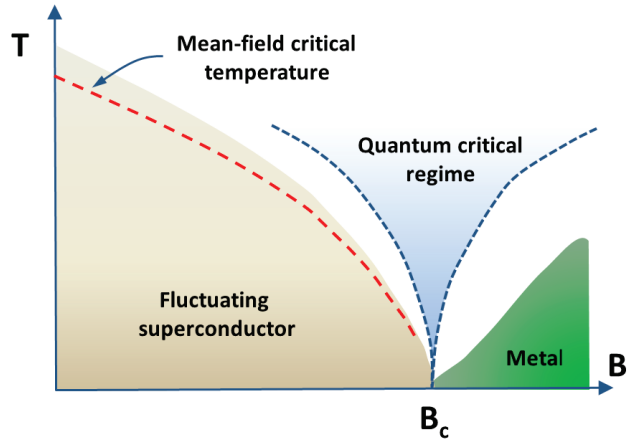

Fig. 1. Schematic phase diagram for the pair-breaking superconductor to metal quantum phase transition in nanowires. The magnetic field dephases the effectively 1d Cooper pairs and suppresses the superconducting critical temperature to zero at a critical field $B_c$. In the quantum critical regime under consideration here, the existence of the quantum critical point results in large corrections to the nanowire conductivity due to superconducting fluctuations.

In the simplest scenario, a quantum critical regime emerging between superconducting and metallic ground states should be controlled by microscopic processes present in the neighboring phases. In this quantum regime, it was proposed that the behavior of superconducting nanowires can be *quantitatively* described by a strongly-coupled pair-breaking quantum field theory[23,24]. This theory captures the universal dynamics of strongly interacting fluctuating 1d Cooper pairs that are unable to form the condensate due to the existence of the magnetic field and that are overdamped (experience an effective frictional force) due to their interactions with a semi-infinite bath of uncondensed bulk electrons in the nanowire. Unlike theories that only capture phase fluctuations of the resulting superconducting order parameter, here amplitude fluctuations are also included in the dynamics. The resulting singular contribution to the conductivity due to superconducting fluctuations is predicted to take the scaling form

$$\sigma(T) = \frac{(e^*)^2}{\hbar}\left(\frac{\hbar D}{k_B T}\right)^{1/z} \Phi_\sigma\left(\frac{\hbar(\alpha-\alpha_C)^\nu}{k_B T^{1/z}}\right), \quad (1)$$

where $e^* = 2e$ is the charge of a Cooper pair, $D$ is the diffusion coefficient, $\Phi_\sigma(x)$ is a dimensionless universal scaling function, $\alpha$ is the pair-breaking frequency and $\alpha_c$ its critical value, $\nu$ is the correlation length exponent and $z$ is the dynamical exponent. The prefactor in Eq. (1) is a product of conductance, $(e^*)^2/\hbar$, and the thermal length, $L \sim T^{-1/z}$, the only available length scale of the problem which describes the size of a superconducting region of the nanowire. The prefactor represents the 1d-case for the dependence $\sigma \sim T^{(d-2)/z}$ introduced from general considerations for the dynamic conductivity in the critical regime.[25,26] The same dependence is obtained from a class of holographic models using gauge-gravity duality.[27] The breakthrough aspect of the field theory that distinguishes it from earlier works on finite-size scaling analysis is that the entire function $\Phi_\sigma(x)$ is theoretically-computed and therefore provides an unequivocal description of the critical regime of the QPT.

To verify the presence of the pair-breaking QPT in 1d superconductors we have studied a magnetic-field driven transition in nanowires made of amorphous Mo-Ge alloys. Two studied nanowires, labeled E and D, had the same length, $L = 3$ μm, but different thickness, $t$, width, $w$, relative content of Mo and Ge, and as a result different $T_c$. (Nanowire E: $Mo_{78}Ge_{22}$, $t = 6$ nm, $w = 13$ nm, $T_c = 1.5$ K; nanowire D: $Mo_{50}Ge_{50}$, $t = 10$ nm, $w = 25$ nm, $T_c = 0.6$ K). Nanowires were fabricated using electron beam lithography with a negative resist; a scanning electron microscopy (SEM) image of nanowire D is shown in Fig.2(a). The method provided excellent uniformity of wires with very small variation of width, ±0.7 nm [Fig.2(b)]. Parameters of Mo-Ge alloys and nanowires are given in the supplementary material (SM). For both nanowires, transport measurements were made in parallel magnetic field and in a field transverse (perpendicular) to the long axis of the wire.

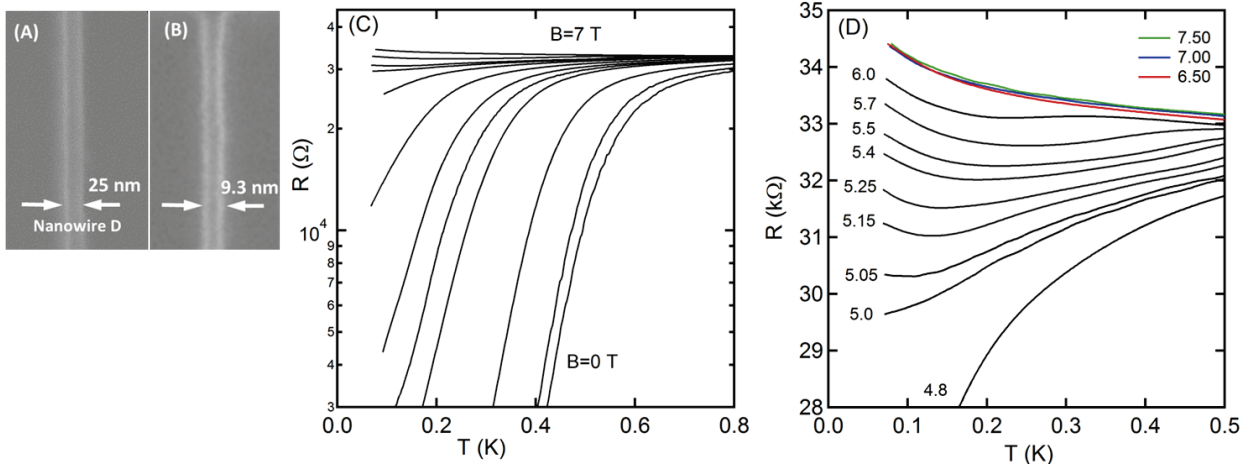

**Fig. 2 Superconductor-metal transition in nanowires.** (A) Scanning electron microscopy (SEM) image of nanowire D. (B) SEM image of one of the thinnest fabricated nanowires (the wire was insulating in zero magnetic field and was not used in the present study), which demonstrates high uniformity of a nanowire's width achieved by the employed method. (C) Resistance versus temperature for nanowire D in parallel magnetic field in the range 0-7.50 T on a logarithmic scale. (D) Same data on a linear scale near the critical field. Notice that the data at the fields 6.50, 7.00 and 7.50 T fall on top of each other indicating the complete suppression of superconducting fluctuations for these fields.

Fig. 2(c) displays the temperature dependence of the resistance, $R(T)$ for nanowire D measured at low-bias in a parallel magnetic field. At high fields [Fig.2(d)], the $R(T)$ dependence first reveals a weak re-entrant behavior and then, for $B \geq 6.5\,\text{T}$, becomes monotonic and field-independent. This behavior indicates that at the fields above 6.50 T the superconducting fluctuations in the wire are completely suppressed. Qualitatively, the same variation was observed in a transverse magnetic field, and for nanowire E. In the high-field regime, the resistance follows the dependence expected for a normal 1d metal, $R_{HF}(T) = R_0 + b/T^\gamma$, which contains the Drude and quantum correction terms. For nanowire E, $\gamma \approx 0.5$, which corresponds to the correction caused by electron-electron interactions; for nanowire D a smaller value, $\gamma \approx 0.27$, was found, likely because this wire is not strictly in the 1d regime for normal electrons ($w < L_T$). At high-biases the wires display a positive zero bias anomaly (not shown) due to electron heating.[28]

The critical behavior described by Eq. (1) is associated exclusively with superconducting fluctuations. However, the nanowire conductance also has a contribution from normal electrons, $G_N(T)$. As a first approximation, we assume that in the critical regime $G_N(T)$ does not change with the field and we take $G_N(T) = 1/R_{HF}(T)$; the conductance of the superconducting channel is then determined as $G_S(T) = 1/R(T,B) - G_N(T)$. Equation (1) indicates that $G_S(B)T^{1/z}$ versus $B$ curves measured at different temperatures should cross at the critical field $B = B_c$. This crossing is indeed observed for nanowire D, as insets in Fig. 3(a,b) show, when we use the value $z \approx 2$ predicted by the critical theory in the "large-N" (N is number of components of the order parameter) limit.[23,24] Looking at Fig.2(d) we notice that at the critical field, $B_c = 5.0$ T, the resistance of the wire is 90% of its value in the normal state (at 7.5 T). This provides a posteriori justification of the approximation used to obtain $G_S(T,B)$. The value of the critical exponent $z \approx 2$ predicted by the pair-breaking critical theory is distinct from $z \approx 1$ value typically associated with the Bose insulator state[13,25] that was observed experimentally in MoGe films[12] and 1d Josephson junction arrays.[29]

    The pair-breaking frequency $\alpha$ in 1d superconductors with strong spin-orbit scattering can be related to magnetic field as $\alpha = kB^2$, where the coefficient $k$ contains both spin and orbital contributions and depends on the orientation of the field (see SM for more details). We use this relation to verify the scaling behavior predicted by Eq. (1) and plotted the quantity $G_S(B,T)T^{1/2}$ versus the scaling parameter $|B^2 - B_c^2|/T^{1/z\nu}$. For both nanowires we obtained $G_S(B)$ from $R(T)$ curves measured at fixed fields. We found that for nanowire D, fixing the correlation length critical exponent to $\nu = 1$ predicted by the pair-breaking theory[21] provides much better data collapse for both field orientations than the non-interacting value, $\nu = 1/2$. Figure 3 displays the results. For nanowire E we did not detect a clear crossing in $G_S(B,T)T^{1/2}$ versus $B$ curves. Nevertheless, we found that at certain value of $B_c$, a fairly good data collapse occurs on the insulating side of the transition (lower branch in Fig.3). We give more comments on the quality of scaling below. The figure also indicates the critical fields for each data set; expectedly, $B_c$ in the parallel orientation is substantially larger than in transverse one. For *both* nanowires the experimental values of $B_c$ are quite close to their values estimated from mean field theory (see SM for details); this indirectly supports our method of finding $B_c$ for nanowire E.

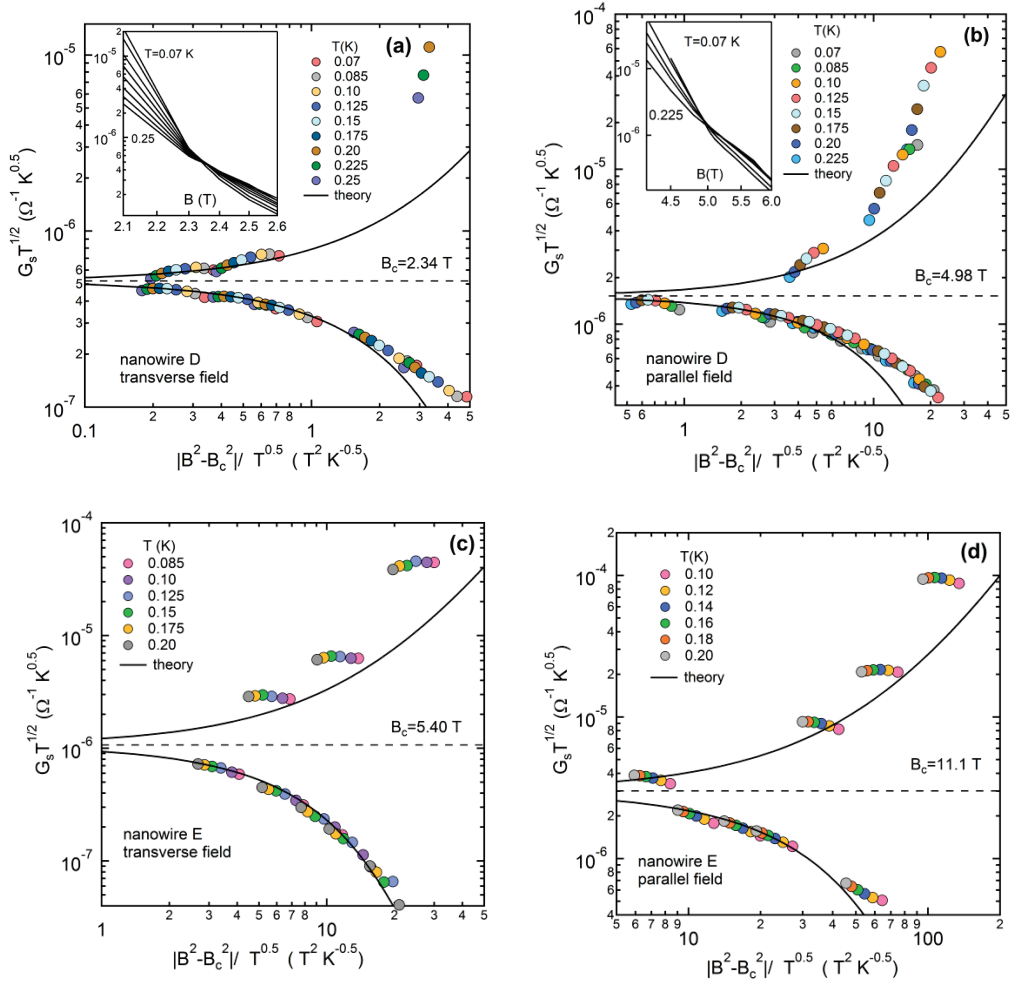

**Fig. 3. Finite-size scaling analysis.** Conductance of the superconducting channel times $T^{1/2}$, $G_S T^{1/2}$, versus the scaling variable $|B^2 - B_c^2|/T^{1/z\nu}$ (with the critical exponents product fixed to be $z\nu = 2$) at different temperatures for nanowires E and D in parallel and transverse orientations of magnetic field. The upper branch corresponds to the superconducting phase ($B < B_c$); the lower to the insulating phase ($B > B_c$). Each panel indicates the critical field, $B_c$. The solid lines indicate the prediction of the pair breaking critical theory. The inset to panel (a,b) shows $G_S T^{1/2}$ versus magnetic field.

The critical exponents are determined by the most general properties of a system, such as dimensionality and the symmetry of the order parameter. They can help to identify a universality class of a QPT, but by themselves do not provide much information about the microscopic physics of the transition. What markedly sets our work apart from the majority of studies of QPTs is the possibility to quantitatively compare experimental data with the critical theory, which predicts not only the exponents but also the scaling function itself, $\Phi_\sigma(x)$. The scaling function was computed numerically in Ref [24]. A brief summary of the theory and the dependence of $\Phi_\sigma(x)$ on its argument are presented in SM. First of all, the theory makes a universal prediction that at the critical field $\Phi_\sigma(x) \approx 0.218$. Using experimental values of $G_S(B = B_c)T^{1/2}$ and estimated diffusion coefficients of $Mo_{78}Ge_{22}$ (0.5 cm$^2$/s) and $Mo_{50}Ge_{50}$ (0.45 cm$^2$/s) alloys (see SM for details), we found that $\Phi_{\sigma\perp}(0) \approx 0.16$ and $\Phi_{\sigma\parallel}(0) \approx 0.46$ for nanowire E and $\Phi_{\sigma\perp}(0) \approx 0.085$ and $\Phi_{\sigma\parallel}(0) \approx 0.24$ for nanowire D, where the sub-index indicates the field orientation. The experiment reproduces the universal number within about a factor of two; this is encouraging given the simple approximation used for the extraction the conduction of the superconducting channel, uncertainty in the diffusion coefficient and the non-trivial prediction that at the critical field $\sigma \sim T^{(d-2)/z}$.

Another advantage of the pair-breaking QPT in 1d superconductors is that the phenomenological coupling parameters of the effective field theory can be connected to those of the microscopic BCS theory. The details are given in the SM; the final result is that the parameter of the scaling function can be written as

$$\Phi_\sigma(x) = \Phi_\sigma \left( C \times 0.54 \left( \frac{\hbar k_B}{D} \right)^{1/2} (k_F \ell)^{1/2} \frac{A\sigma_{3d} T_c}{e^2} \frac{(B^2 - B_c^2)}{B_c^2 T^{1/2}} \right), \quad (2)$$

Here $T_c$, $B_c$, $D$, cross section area $A$, and bulk conductivity $\sigma_{3d}$ are known nanowire parameters. The mean free path, $\ell$, in amorphous Mo-Ge alloys is roughly equal to interatomic distance, 0.3 nm; this assumption allows us to estimate the Fermi vector, $k_F$. The dimensionless constant $C$ connects the bare and renormalized pair-breaking strengths and is the only adjustable parameter. Equation 2 allows us to test the analytical form of the scaling function and the relation of its argument to non-universal parameters characterizing a nanowire.

To make comparison with the theory, in Fig.4 (a,b) we plot the quantity $G_S T^{1/2}$ normalized by its value at the critical field versus the normalized scaling variable $|B^2 - B_c^2|/T^{1/2}B_c^2$. As predicted by Eq. 2 after normalization, the data for two field orientations coincide. The scaling function $\Phi_\sigma(x)$ is known from the theory in numerical form; we plotted it in the figure as $\Phi_\sigma(x)/\Phi_\sigma(0)$ and adjusted constant $C$ to fit the data. Remarkably we found that for nanowire E, with $C \approx 0.05$ the theory matches the *non-linear* variation of the data in the insulating regime (bottom branch in panel (b)). For nanowire D, the close value of the constant $C \approx 0.04$ also gives a good fit for both superconducting and insulating branches of the data, albeit in more narrow range of fields near $B_c$. The relatively small value of $C$ could be indicative of possible deviations of the pair-breaking frequency from its mean-field value due to strong fluctuations near the QPT and also a multiplicative effect of small errors in microscopic parameters used in the scaling function argument. For completeness we plotted the theoretical predictions for the quantity $G_S T^{1/2}$ on top of each data set in Fig. 3.

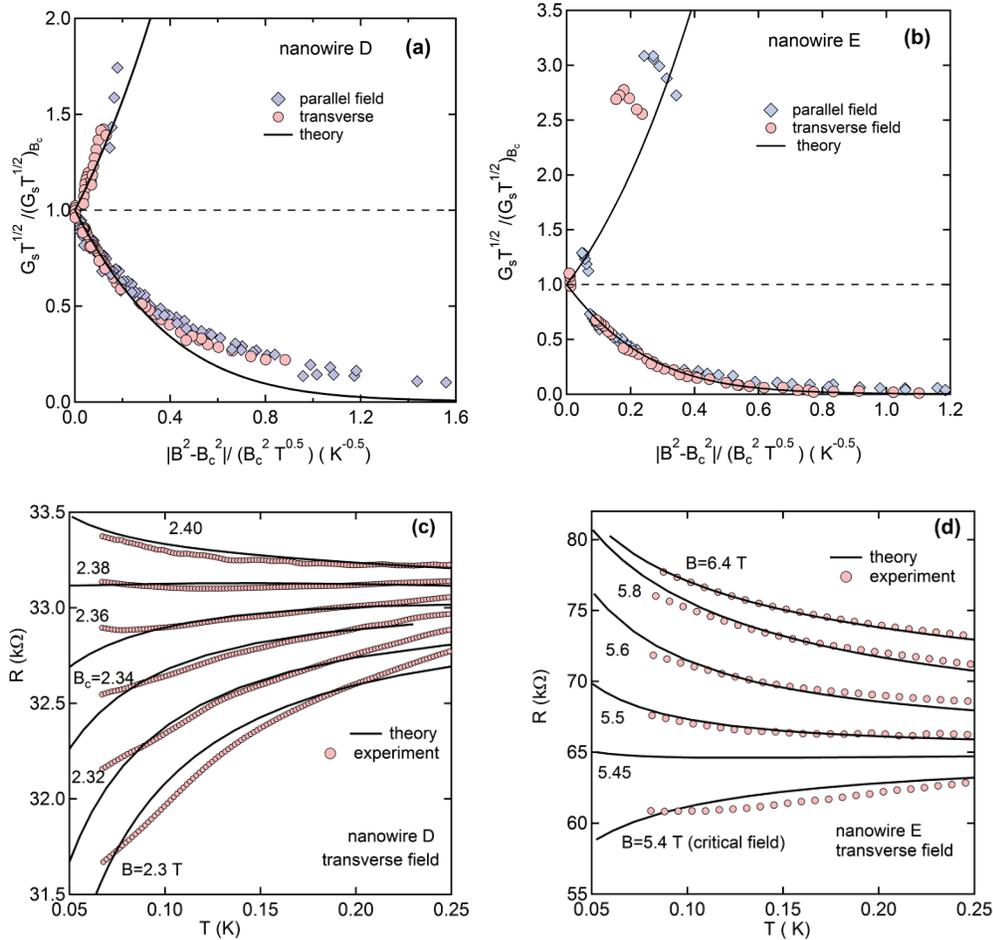

**Fig. 4. Quantitative test of the pair-breaking QPT theory.** (a,b) Conductance of superconducting channel times $T^{1/2}$, $G_S T^{1/2}$, normalized by its value at the critical field versus normalized scaling variable $\left| B^2 - B_c^2 \right| / T^{1/2} B_c^2$ for nanowires D (panel a) and E (panel b) in parallel and transverse field orientations. Black solid curves are predictions for the full scaling function in the quantum critical regime computed from the pair-breaking critical theory. (c) Experimental $R(T)$ curves superimposed with $R(T)$ curves computed from the critical theory across the QPT for nanowire D in transverse field. (d) Same data for nanowire E in the transverse field in the insulating regime.

With all parameters of the theory determined from the fit to the scaling function, we now reverse our analysis, and for each nanowire and field orientation use Eq. 1, Eq. 2 and experimental conductance of the normal electrons, $G_N(T) = 1/R_{HF}(T)$, to compute theoretical $R_{th}(T)$ dependences corresponding to the series of the experimental $R(T)$ curves near the transition. The theoretical and experimental $R(T)$ dependences are shown in Fig. 4 (c,d). We emphasize that all theory curves were generated with no additional adjustment; we used the values of $C$ obtained from the scaling fit shown in Fig. 4(a,b). So the overall agreement with the theory is remarkable. In Fig. 4(d) we added theoretical $R_{th}(T)$ for $B = 5.45$ T, which indicates that the weak reentrant behavior, at least in principle, can be captured by the theory (more details are given in SM).

We further notice that the deviations from the theory for all data shown in Figs. 3 and 4 display a common trend. The agreement with the theory and the quality of the data scaling are the best in the insulating regime; however they both becomes worse near the critical field and then even more so in the superconducting regime. The most likely reason is that with the evolution of a nanowire across $B_c$ our approximation for the conduction of the normal channel, which is valid only in the perturbative limit, becomes progressively less accurate. Indeed, once superconducting fluctuations develop, the time-average density of normal electrons decreases. This should not only increase the temperature-independent Drude term in $R_N(T)$ but also alter its temperature dependence due to the modification of the quantum corrections. Apparently these effects are more significant in nanowire E in which we did not observe the scaling of the conductance in the superconducting regime [Fig 3(c,d)]; only a general variation of $G_S T^{1/2}$ with the field follows the theory. This wire has roughly three times smaller cross sectional area than nanowire D and more pronounced $R_{HF}(T)$ dependence.

The agreement with the critical theory establishes that a quantum phase transition indeed takes place in 1d superconductors; this was not an obvious scenario to start with. Moreover, the fact that for both nanowires and field orientations the experiment reproduces the non-trivial dependence of the scaling function argument on several non-universal nanowire parameters, confirms that the theory accurately captures the microscopic physics at the transition. This allows us to draw several conclusions. The values of exponents, $\nu$ and $z$, indicate strong interactions between superconducting fluctuations mediated by the normal electrons and validate the "large-N" approximation used in the theory. This tells us that in the quantum critical regime the nanowire dynamically splits into semi-independent segments due to both phase and amplitude fluctuations that locally suppress the order parameter. The value of $z \approx 2$ and overall agreement with the pair-breaking critical theory allows us to rule out the unbinding of quantum phase-slip and anti-phase slip pairs [16] as an alternative mechanism of the QPT. This later transition belongs to the "phase-only" Berezinskii-Kosterlitz-Thouless (BKT) class and is claimed to occur in 1d Josephson-Junction arrays [29]. A phase slip temporarily brings to a normal state a section of wire, which globally is in the superconducting state. In our case, in the quantum critical region shown in Fig. 1, both amplitude and phase fluctuations of the superconducting order parameters must be treated on equal footing due to the presence of the pair-breaking energy scale. The critical fluctuations responsible for the magnetic-field driven QPT in nanowires microscopically are not phase slips, but Aslamazov-Larkin (AL) type fluctuations [22,23,24], which produce superconducting regions in a normal metal. However, the AL theory must be modified in $d = 1$ to take into account the interactions between Cooper pairs overdamped due to their coupling to the fermionic quasiparticles of the proximate superconductor.

The obtained value of the critical exponent, $\nu \approx 1$, is in violation of the so-called Harris criteria for a disordered system, $\nu \geq 2/d$; *i.e.*, the system is effectively in a clean limit for the temperature range we consider where disorder acts on scales smaller than the thermal length (see SM for more details).

Our findings are in accord with an earlier conjecture about the presence of dissipative effects in disordered superconducting films near the critical point.[30] A similarity between wires and films is expected since within perturbation theory[22] the microscopic effect of a magnetic field on both systems is the same. This suggests that a critical pair-breaking theory incorporating physical processes found to be relevant for nanowires may provide a not-yet-known microscopic description of QPTs in 2d disordered films of conventional superconductors. This approach can also be extended to describe a disorder-driven QPT in anisotropic gap superconductors where non-magnetic disorder acts as a pair-breaker. In fact, a theory[31] describing the behavior of the magnetic susceptibility near a pair-breaking QPT in 2d films has been developed recently; a computation of conductivity within this theory is highly desirable for comparison with experiments. Future work can also explore a predicted correspondence between QPTs in superconducting nanowires and films and magnetic systems.[31,32]

In summary, the excellent agreement with the quantum critical theory observed across the superconductor-meal transition in MoGe nanowires supports its microscopic underpinnings and represents an important benchmark in the confirmation of universality at quantum phase transitions.

**Methods:**

**1. Nanowire fabrication**

The nanowires were fabricated using Si wafers covered with a 100 nm layer of SiN and cut in individual samples with size 6x9 mm$^2$. First, using optical photolithography, consequential deposition of Ti (20 nm) and Au (40 nm) films, and liftoff procedure we fabricated a pattern consisting of 12 electrodes and several markers. Next, we sputter deposited a layer of amorphous Ge (thickness 3 nm) followed (without breaking vacuum) by the sputter deposition of MoGe alloy. To make good electrical connection between pre-patterned Ti/Au electrodes and thin MoGe films, square pads (5x5 micrometers, thickness 30 nm) were placed in each contact area by positive electron beam lithography with PMAA resist and liftoff. After patterning contact pads, the sample was immersed in the 2.5 % water solution of TMAH (the developer for negative electron beam lithography) for clearing. In the next step the whole sample was spin coated with 35-nm thick HSQ (hydrogen silsequioxane) layer. The specification of the solution is XR-1541 2 %; it was purchased from Dow Corning. The nanowire and film electrodes were patterned by electron-beam lithography in Nova Nano 630 Scanning Electron Microscope. The accelerating voltage was 30 keV; the dosage was 400-600 μC/cm$^2$ for electrode areas and 3-8 nC/cm for nanowire lines. The exposed pattern was developed in 2.5 % water solution of TMAH for 2 min to remove HSQ. The pattern was etched with reactive ion etching using SF$_6$ gas.

**2. Transport measurements**

Measurements were carried out in a dilution refrigerator equipped with a superconducting solenoid. The use of lossy miniature stainless steel coaxial cables, room temperature feedthrough filters, and capacitance to ground mounted directly on the sample holder at low temperature enable to preclude spurious saturation of low resistive states at the lowest temperature. All measurements were performed with lock-in amplifier technics and high input impedance voltage amplifiers in current bias configuration (~0.1 – 0.5 nA).

## 3. Data Analysis

The analysis employed the scaling function computed in the theory [24] in a numerical form. The microscopic parameters of Mo-Ge alloys were computed from the known (or approximated) conductivity, specific heat and mass density data. The details and a brief summary of the theory are given in SM.

**Acknowledgments:** The authors thank N. Shah, B, Rosenow, S. Sachdev and O. Starykh for valuable comments. Nanowire fabrication was carried out at the University of Utah Microfab and USTAR facilities. A.R. acknowledges Université Grenoble Alpes and Institute Néel where measurements were performed, for hospitality. This research was supported in part by the National Science Foundation under Awards No. DMR-1611421 (A.R.) and DMR-1553991 (A.D.) and by the ERC Grant QUEST #637815 (B.S.). A.D. performed a portion of this work at the Aspen Center for Physics, which is supported by NSF grant PHY-1607611.

**Author Contributions**: H.K. fabricated nanowires, A.R., B.S. and F.G. carried out measurements, A.R., B.S., and A.D. carried out data analysis and wrote the manuscript, A.R. conceived and coordinated the project.

# Supplementary Materials:

## 1. Parameters of $Mo_{78}Ge_{22}$ and $Mo_{50}Ge_{50}$ amorphous alloys

Mo-Ge alloys have an amorphous atomic structure; a mean free path in these materials is close to the interatomic distance. As a result the bulk resistivity, $\rho_v$, at room temperature (where quantum correction can be neglected) does not depend on size of a nanowire. It depends on relative content of Mo and Ge: $Mo_{78}Ge_{22}$ ($\rho_v = 160$ μΩ cm) and $Mo_{50}Ge_{50}$ ($\rho_v = 235$ μΩ cm). The cross sectional area, $A$, of a nanowire can be estimated from room temperature resistance, $R_0$, as $A = \rho_v L / R_0$.

For the analysis of the data presented in the main part of the paper we need to summarize and estimate some physical parameters of the two amorphous Mo-Ge alloys. To estimate the mass density of amorphous $Mo_{78}Ge_{22}$ alloy we use the known density of intermetallic compound $Mo_3Ge$, $\rho_c = 9.97$ g/cm$^3$, [S1] and assumed that it is reduced by a factor 0.86=0.64/0.74 (ratio of filling factors for randomly and closely-packed spheres). This gives $\rho_a = 8.60$ g/cm$^3$. The density of $a$-$Mo_{50}Ge_{50}$ is estimated using the same factor 0.86 and extrapolation between the densities of two intermetallic compounds $Mo_5Ge_3$ ($\rho_c = 9.63$ g/cm$^3$ [S1]) and $MoGe_2$ ($\rho_c = 8.83$ g/cm$^3$ [S1]) ; this gives $\rho_a = 8.0$ g/cm$^3$. The electron specific heat coefficient $\gamma$ is 3.05 mJ/mole K$^2$ for $Mo_{78}Ge_{22}$ and 2.32 mJ/mole K$^2$ for $Mo_{50}Ge_{50}$ (extrapolated between two neighboring data points) [S2]. Using $\rho_a$ and $\gamma$, we estimate electronic specific heat per unit volume, $\gamma_V = 2.9 \times 10^2$ J/m$^3$K$^2$ for $a$-$Mo_{78}Ge_{22}$ and $\gamma_V = 2.2 \times 10^2$ J/m$^3$K$^2$ for $a$-$Mo_{50}Ge_{50}$ and then the density of states at the Fermi level, $g(0)$, using relation $\gamma_V = \pi^2 k_B^2 g(0)/3$; numerical values of $g(0)$ are given in Table S1. These estimates ignore possible electron-phonon enhancement of $\gamma$ by the factor of 1.1-2 [S3] To proceed further we assume the Fermi surface of the alloys is spherical, but because of the presence of a d-element the effective mass of carriers, $m^*$, is not equal to the free electron mass $m_e$; this is fairly standard starting approximation for amorphous or liquid metals [S4]. We also assume following Ref. [S5] that the mean free path is equal to interatomic distance $\ell \approx 0.3$ nm. This is just a good guess; unfortunately because of the unknown $m^*$, $\ell$ can't be determined accurately. The choice of $\ell$ is supported by the experimentally determined value of $\ell \approx 0.43$ nm in amorphous $Ag_{40}Cu_{40}Ge_{20}$ alloy, which, according to the Hall and specific heat measurements, follows accurately the free electron model with the electron density consistent with one free electron provided by each Ag and Cu atom and four electrons by Ge [S6]. Somewhat smaller value of $\ell$ is expected in amorphous alloy with d-electrons. Taking $\ell \approx 0.3$ nm and using the free electron model relations $\sigma = e^2 n\tau / m^* = e^2 \ell k_F^2 / 3\pi^2 \hbar = e^2 D g(0)$ and $n = k_F^3 / 3\pi^2$ we find parameters of the alloys listed in the table.

The spin-orbit scattering time in the alloys can be estimated as $\tau_{so} = \tau \left( \hbar c 4\pi\varepsilon_0 / Ze^2 \right)^4$, where $\varepsilon_0$ is the permittivity of free space and $Z$ is the atomic number of an element. For an alloy with composition $Mo_xGe_{1-x}$, we estimate $Z$ as $Z = xZ_{Mo} + (1-x)Z_{Ge}$; numerical values of $\tau_{so}$ are listed in the table. These estimates provided a good agreement with the mean-field pair-breaking critical field in MoGe and Nb nanowires [S7]. It also gives the spin-orbit scattering length for MoGe alloys, $\ell_{so} = \sqrt{D\tau_{so}} \approx 1.5$ nm, consistent with the boundary between two regimes in MoGe films doped with Gd [S8] (Our estimate has two orders of magnitude disagreement with the value $\tau_{so} \simeq 1.3 \times 10^{-12}$ s obtained from weak localization correction measurements [S5]. This number is clearly wrong since it would give the critical field an order of magnitude smaller than experimental value).

|  | $Mo_{78}Ge_{22}$ | $Mo_{50}Ge_{50}$ |
|---|---|---|
| Room temperature resistivity, $\rho_v$ ($\mu\Omega$ cm) | 160 | 235 |
| Mass density, $\rho_a$ (g/cm$^3$) | 8.60 | 8.0 |
| Density of states at the Fermi level, $g(0)$ (J$^{-1}$m$^{-3}$) | 4.6 x 10$^{47}$ | 3.5 x 10$^{47}$ |
| Diffusion coefficient, $D$ (m$^2$/s) | 5.3 x 10$^{-5}$ | 4.7 x 10$^{-5}$ |
| Carrier density, $n$ (m$^{-3}$) | 1.4 x 10$^{29}$ | 0.75 x 10$^{29}$ |
| Fermi wave vector, $k_F$ (m$^{-1}$) | 1.6 x 10$^{10}$ | 1.3 x 10$^{10}$ |
| Elastic scattering time, $\tau$ (s) | 5.6 x 10$^{-16}$ | 6.5 x 10$^{-16}$ |
| Effective carrier mass $m^*/m_e$ | 3.5 | 3.2 |
| Spin orbit scattering time, $\tau_{so}$ (s) | 7.6 x 10$^{-14}$ | 1.2 x 10$^{-13}$ |

**Table S1**. Parameters of Mo-Ge amorphous alloys.

**2. Parameters of nanowires E and D. Estimate of critical magnetic field from mean-field pair-breaking theory.**

Parameters on nanowires, E and D, studied in this work are listed in the Table S2. A wire is classified as one-dimensional superconductor if its width and thickness are smaller than $w, t < \pi\sqrt{2}\xi(0)$ [S9]. This corresponds to the case when formation of vortices in a wire is not energetically favorable [S10]. The Ginzburg-Landau coherence length can be estimated as $\xi(0) = 0.85(\ell\xi_0)^{1/2}$, where the clean limit coherence length can be determined as $\xi_0 = \hbar v_F / 1.764\pi k_B T_C$. Using the value of mean free path $\ell \approx 0.3$ nm and other parameters listed in Tables 1 and 2, we find $\xi(0) \approx 10$ nm for nanowire E and $\xi(0) \approx 15$ nm for nanowire D. So both wires fall within 1D superconducting limit.

|  | E (pl) | E (tr) | D (pl) | D (tr) |
|---|---|---|---|---|
|  | $Mo_{78}Ge_{22}$ | $Mo_{78}Ge_{22}$ | $Mo_{50}Ge_{50}$ | $Mo_{50}Ge_{50}$ |
| $L$ ($\mu$m) | 3 | 3 | 3 | 3 |
| $t$ (nm) | 6 | 6 | 10 | 10 |
| $w$ (nm) | 13 | 13 | 25 | 25 |
| $T_C$ (K) | 1.5 | 1.5 | 0.6 | 0.6 |
| $R_{RT}$ (k$\Omega$) | 62 | 62 | 27 | 27 |
| $B_C$ (T) | 11.1 | 5.40 | 4.98 | 2.34 |
| $B_{MF}$ (T) | 14.2 | 7.0 | 5.4 | 2.4 |
| $\xi(0)$ (nm) | 10 | 10 | 15 | 15 |
| $(G_S\sqrt{T})_C$ ($\Omega^{-1}K^{1/2}$) | $2.8 \times 10^{-6}$ | $1.1 \times 10^{-6}$ | $1.55 \times 10^{-6}$ | $0.5 \times 10^{-6}$ |
| $\Phi_\sigma(0)_{EXP}$ | 0.46 | 0.16 | 0.24 | 0.085 |

**Table S2**. Parameters of nanowires E and D in parallel (pl) and transverse (tr) magnetic fields. $T_C$ mean-field critical temperature, $B_C$ experimental critical field, $B_{MF}$ mean field critical field estimated using nanowire geometry and microscopic parameters of the alloys, $\xi(0)$ -Ginzburg-Landau coherence length at zero temperature. Quantity $(G_S\sqrt{T})_C$ gives the value of the $G_S(T)\sqrt{T}$ at the critical field, where $G_S(T)$ is the conductance of superconducting channel of a wire. $\Phi_\sigma(0)_{EXP}$ is experimentally determined value of the scaling function (see Eq. 1) at the critical field. The theory predicts that it assumes universal value $\Phi_\sigma(0) \approx 0.218$.

Within mean-field approximation the magnetic field suppression of superconductivity in disordered nanowires with sufficiently strong spin-orbit scattering comes from spin and orbital pair-breakers. The strength of the pair-breaker acting on the spin degree of freedom is given as $\tilde{\alpha}_s \approx \hbar \tau_{so} e^2 B^2 / 2m^2$ [S11] To estimates its magnitude in MoGe alloys we have used effective mass $m = m^*$ and spin orbit scattering time $\tau_{so}$ listed in Tab. 1. The orbital pair-breaker can be found from relation $\tilde{\alpha}_o = 2De^2 \langle A^2 \rangle / \hbar$, [S11] where $\langle A^2 \rangle$ is averaged value of the squared vector potential in a wire. This relation also can be applied for thin films in parallel magnetic field and for small particles. For 1d-regime it is assumed that screening is negligible and magnetic field in a wire is uniform and equal to the external field $B$. For field orientation transverse to the wire width one can choose $\vec{A} = (0, Bx, 0)$, which after averaging gives $\langle A^2 \rangle = B^2 w^2 / 12$. For nanowire in parallel field orientations we approximate our nanowire as a cylinder with cross sectional area equal to that of the wire. The effective radius of the wire is $R = \sqrt{wt/\pi}$. Then, using the gauge $\vec{A} = (\vec{B} \times \vec{r})/2$, we get $\langle A^2 \rangle = B^2 R^2 / 8 = B^2 wt / 8\pi$. We notice that both spin and orbital pair-breakers are proportional to $B^2$; total pair-breaking strength is the sum of two contributions, $\tilde{\alpha} = \tilde{\alpha}_s + \tilde{\alpha}_o$. The critical value of the pair-braking strength can be estimated within mean-field theory as $2\tilde{\alpha}_c = 1.76 k_B T_c$. The estimated values of mean-field critical field are listed in Tab.S2. They are somewhat sensitive to not-exactly-known effective mass, $m^*$, but overall are close to the critical fields detected experimentally.

**3. Dimensionless scaling at the superconductor to metal transition in nanowires.**

In this section, we reproduce some of the relevant details from Refs. [S12-S13] which allow us to connect the dimensionless argument of the scaling function in Eq. 1 with the microscopic parameters of the experiment as reflected in Eq. 2 of the main text.

**3.1 Pair-Breaking Theory**

The starting point is an effective theory of repulsive Cooper pairs in one spatial dimension without charge conservation in the condensate due to the existence of a large bath of unpaired fermions (transverse conduction channels) in the metallic nanowire. The resulting quantum critical action for the ohmically damped complex Cooper pair order parameter $\Psi(z,\tau)$ is:

$$S = \int_0^L dz \int_0^{\hbar\beta} d\tau \left[ D |\partial_z \Psi(z,\tau)|^2 + \alpha |\Psi(z,\tau)|^2 + \frac{u}{2} |\Psi(z,\tau)|^4 \right] \\ + \frac{\eta}{\hbar\beta} \sum_{\omega_n} \int_0^L dz \, |\omega_n| |\Psi(z,\omega_n)|^2 \tag{S1}$$

where $z$ is a coordinate along the wire, $\beta = 1/k_B T$, $\omega_n$ a bosonic Matsubara frequency, $D = v_F \ell/3$ is the diffusion constant with $v_F$ the Fermi velocity and $\ell$ is the mean free path. Interactions are parametrized by $u > 0$ which is relevenat in $d = 1$ and a quantum phase transition at $\alpha_c$ can be tuned by altering the strength of the pair-breaking frequency $\alpha$ which can be connected to the physical magnetic field (see below). The dynamics of the Cooper pairs are subject to damping due to their decay into the metallic bath characterized by the parameter $\eta$ and we have defined the Fourier transform in imaginary time as:

$$\Psi(z,\omega_n) = \int_0^{\hbar\beta} d\tau \, \Psi(z,\tau) e^{i\omega_n \tau}. \tag{S2}$$

Simple power counting at tree level gives the bare value of the dynamical critical exponent to be $z = 2$.

### 3.2 Effects of Disorder

In principle, both the diffusion constant and pair breaking in Eq. (S1) can be random functions of position $x$ along the wire. Disorder in $\alpha$ is related to fluctuations in the local density of states and is expected to be relevant at $T = 0$ at the quantum-critical point as defined by the Harris criterion since $1 \approx d\nu < 2$. A systematic study of the strongly disordered theory is presented in Ref. [S14] but here we neglect this randomness, and are thus restricted to the quantum critical regime at temperatures above $T_{dis}$ determined by equating the thermal length $L_T = \sqrt{\hbar D/(k_B T_{dis})}$ to a zero-magnetic field disorder length scale $\ell_{dis} \equiv \ell N_\perp$ where $N_\perp = 2 n A/k_F$ is the number of transverse metallic conduction channels and $A = t\,w$ is the cross-sectional area of the nanowire [S15]. This sets:

$$T_{dis} = \frac{\hbar}{3 k_B N_\perp^2 \tau} \tag{S3}$$

where $\tau = \ell/v_F$ is the elastic scattering time given in Table S1. Using estimates for non-interacting electrons we find $T_{dis} \approx 0.5 - 2.5$ mK which is far below the base temperature of our experiment confirming the irrelevance of disorder in our results.

### 3.3 Connection with microscopic parameters.

As described in the main text, the nanowires have a background metallic conduction for $\alpha > \alpha_c$ due to $N_\perp$ channels which can be identified from the large field regime due to the complete suppression of superconducting fluctuations. An analysis of Eq. (S1) predicts that all important couplings between bosons and fermions scale to universal values and the singular contribution to the conductance obeys the scaling form given in Eq. (1) of the main text (see Refs. [S11-S12] for details). In the so-called "large-N" approximation where the number of complex components of $\Psi$ is assumed to be large while $z = 2$ and $\nu = 1$ take their bare values, $\Phi_\sigma(x)$ can be evaluated numerically where $x$ is a dimensionless scaling variable:

$$x = C \left(\frac{\hbar D}{k_B T \eta}\right)^{1/2} \frac{\alpha - \alpha_c}{\hbar u} \tag{S4}$$

and $C$ is a non-universal constant related to a renormalization due to the interactions between Cooper pairs.

The critical pair-breaking frequency is related to the critical temperature and critical parallel and transverse magnetic fields for a particular wire as $\alpha_c = 0.88 k T_c / \hbar = \gamma_\| B_{c\|}^2 = \gamma_\perp B_{c\perp}^2$, where $\gamma$ depends on orientation and also takes care of the spin pair-breaking [16-17]. Using this we can re-write

$$\hbar(\alpha - \alpha_c) = \hbar \alpha_c \frac{\alpha - \alpha_c}{\alpha_c} \simeq 0.88 \, k_B T_C \frac{(B^2 - B_c^2)}{B_c^2} \tag{S5}$$

The remaining microscopic parameters $u$ and $\eta$ can be found from the following relations of BCS and time-dependent Ginzburg-Landau theory [S18]:

$$\eta = \frac{\pi^2 \hbar^2}{2 m \, \xi^2(0)} \frac{1}{8 \, k_B T_c} \simeq \frac{1.5}{k_F \ell} \tag{S6a}$$

$$b = \frac{\hbar^2}{m \, \xi^2(0) n} \frac{1}{\chi(0.882 \xi_0/\ell)} \tag{S6b}$$

$$u = \frac{b}{A \, \hbar^2 \eta^2} \simeq 1.33 \, \frac{e^2 D}{\sigma_{3D} A \, \hbar^2} \tag{S6c}$$

where $\xi(0) = 0.85 \sqrt{\xi_0 \ell}$ is the zero-temperature Ginzburg-Landau coherence length, $\xi_0 = 0.18 \, \hbar v_F / k_B T_c$ is the Pippard coherence length and $\chi(z)$ is the Gor'kov function. In the dirty limit $\chi(0.882 \, \xi_0/\ell) = 1.33 \, \ell/\xi_0$. Using the standard equations for the Drude conductivity and combining all terms we find

$$\Phi_\sigma(x) = \Phi_\sigma\left(C \times 0.54 \left(\frac{\hbar k_B}{D}\right)^{1/2} (k_F \ell)^{1/2} \frac{A \sigma_{3d} T_c}{e^2} \frac{(B^2 - B_c^2)}{B_c^2 T^{1/2}}\right), \tag{S7}$$

which appears in Eq. 2 of the main text. The non-universal microscopic constant $C$ is the only adjustable parameter in the theory and it can be extracted by fitting the scaling function to experimental data in the quantum critical regime. This parameter reflects not only the renormalization in the critical pair breaking frequency $\alpha_c$ due to interactions between Cooper pairs, but also the approximations used when matching the coupling constants in the effective field theory to microscopic values via the time-dependent Ginzburg-Landau theory.

Let us notice that whereas the scaling function $\Phi_\sigma(x)$ is monotonically decreasing function of its argument (we reproduce $\Phi_\sigma(x)$ in Fig.1 (a) below), the theoretical dependence of conductance in the insulating regime can behave differently because of the temperature-dependent $T^{-1/2}$ prefactor. In Fig. 1(b) we show several theoretical $G_{S,theory}(T)$ dependences for nanowire E in the transverse field computed using theoretical scaling function with parameters determined from the scaling fits [Fig. 4(b) of the main manuscript]. One can see that depending on the field, the conductance can decrease with temperature ($B = 5.45$ T), display non-monotonic variation ($B = 5.6$ T) or increase ($B = 5.8$ T). The theoretically expected resistance can be computed as $R_{th}(T) = 1/(G_N + G_{S,theory})$, where $G_N = G_{HF}$ is the experimental conductance of the normal channel approximated by the conductance of the nanowire at high magnetic fields. In Fig. 1(c) we show a zoom-in dependence of a theoretical curve at $B = 5.45$ T which shows a weak re-entrant behavior thus confirming that at least in principle this behavior can be reproduced by the theory.

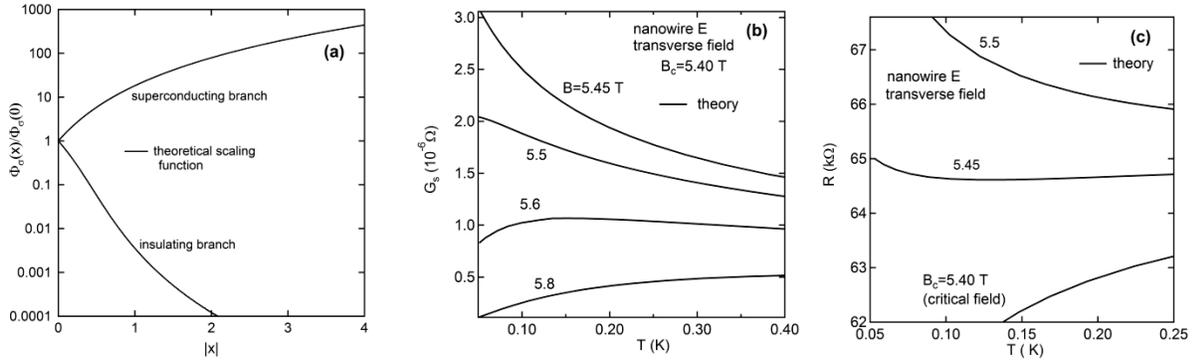

*Fig. 1. (a) The theoretical scaling function versus the absolute value of its argument. (b) Theoretical conductance caused by the critical superconducting fluctuations in the insulating regime of nanowire E at indicated magnetic fields. (c) Resistance of the nanowire E in transverse magnetic field expected from the theory. At $B = 5.45$ T the re-entrant behavior is expected.*